# Group Anonymity

Oleg Chertov[1], Dan Tavrov[1]

[1] Faculty of Applied Mathematics, National Technical University of Ukraine
"Kyiv Polytechnic Institute", 37 Peremohy Prospekt, 03056 Kyiv, Ukraine
{chertov, kmudisco}@i.ua

**Abstract.** In recent years the amount of digital data in the world has risen immensely. But, the more information exists, the greater is the possibility of its unwanted disclosure. Thus, the data privacy protection has become a pressing problem of the present time.
The task of individual privacy-preserving is being thoroughly studied nowadays. At the same time, the problem of statistical disclosure control for collective (or group) data is still open.
In this paper we propose an effective and relatively simple (wavelet-based) way to provide group anonymity in collective data. We also provide a real-life example to illustrate the method.

## 1 Introduction

Year by year more and more data emerge in the world. According to the latest IDC research [1], the Digital Universe will double every 18 months, and the number of "security-intensive" information will grow from 30% to roughly 45% by the end of 2012. This means that the risk of privacy violation and confidential data disclosure rises dramatically. The threat doesn't only comprise the possibility of stealing a credit card and social security numbers, patient medical records and images, Internet commerce and other transaction data. It is popular nowadays to provide direct access to the depersonalized and non-aggregated primary data. E.g., one can easily gain access to the microfile data in census statistics and sociology. But, if these data aren't protected, individual anonymity can easily be violated. That has been clearly shown by Sweeney [2, p. 21]. Using Massachusetts voters data, she proved that knowing a person's birth date and full 9-digit ZIP code is enough to identify 97% of voters.

A problem of individual anonymity in primary data isn't a new one. It is being more or less successfully solved as one of the main tasks in privacy-preserving data mining. There are different statistical disclosure control methods [3] which guarantee:

- $k$-anonymity [4]. This means every attribute values combination corresponds to at least $k$ respondents existing in the dataset sharing the combination;
- even more delicate $l$-diversity [5] and $t$-closeness [6].

At the same time a problem of providing group (collective) anonymity is still open [7]. By *group anonymity* we understand protecting such data features that can't be distinguished by considering individual records only. E.g., we cannot protect the regional distribution of young unemployed females in terms of inidividual anonymity.

Either individual or group anonymity can be provided by introducing an acceptable level of uncertainty to the primary data. By making specific data records impossible to distinguish among the others, we guarantee required privacy-preserving.

When providing data anonymity, both group and individual, it is important to take into account the adversarial model which reflects what information is known to an adversary, and what is not.

In our work, we suppose that the potential adversary doesn't possess any other additional information except for the one contained in the primary data.

In general, there exist a variety of approaches to solving the group anonymity problem. In this paper, we will discuss a so-called extremum transition approach. Its main idea is to swap records with specific attribute values between the positions of their extreme concentrations and the other permitted ones.

Depending on the task definition, we can implement this approach by:
- swapping values of required attributes between respondents;
- transferring a respondent to be protected to the other place of living (to the other place of work). In most cases it is natural to transfer not only a single respondent, but the whole respondent's family as well;
- mere modifying the attribute values.

Of course, it's easy to provide group anonymity. All we need is a permission to move respondents between any possible places (as long as the population number on a particular territory remains stable). But such primary data deformation almost inevitably leads to considerable utility loss. Imagine that we want to transfer some respondents to a particular territory. But, there are not enough people of the same sex and age to fit our new "migrants". Obviously, such a transfer cannot be acceptable.

All this leads to a question. If we know what to modify to provide data group anonymity, what should we preserve to prevent data utility loss?

In general, it is possible to preserve *absolute* quantities only for a *particular* population category (all the population provided, respondents with required values of a certain attribute etc.). But, in many cases researches can be interested in the *relative* values rather than in the absolute ones. Let us consider some typical examples.

1. True quantity of military (or special service) officers can be absolutely confidential. This is also the case for their regional distribution. At the same time, information on their distribution by age, marital status or, say, wife's occupation can be very interesting and useful for sociologists.

2. In developing countries, there is usually no public statistics on a company's income. In this case, information on the company's income growth rates can serve as an important marker of its economic status and development prospective.

We come to conclusion that we need to preserve relations between strata, distribution levels, data ranges rather than the absolute values. But, it's not easy to alter data records with a particular attribute values combination and preserve proportional relations between all the other possible ones. Such a problem seems to be as complex as the $k$-anonymization problem. The latter, as stated in [8], is an NP-hard problem.

Certainly, there are different techniques that can aid in finding a balance between altering primary data and preventing utility loss. For instance, we can try to perform such data swapping that main statistical features such as data mean value and their standard deviation will persist. For example, in [11], a specific normalizing process to preserve these features is introduced.

But, in the current paper we propose to use wavelet transform (WT) instead. Surely, WT doesn't guarantee the persistance of all statistical data features (except for their mean value which will be discussed later in the paper), but it can preserve some information that can come in handy for specific studies.

Generally speaking, WT is an effective way to present a square-integrable signal by a basis obtained from certain wavelet and scaling functions providing its both time and frequency representation. We consider WT to be acceptable because:

- It splits primary data into approximation and multilevel details. To protect data, we can redistribute approximation values considering particular attribute values combinations. Besides, we can prevent utility loss by leaving details unchanged (or by altering them proportionally). In this case, proportional relations between different attribute values ranges will be preserved. To illustrate that, let's refer to [9]. In Russia, studying the responses to 44 public opinion polls (1994-2001) showed the following result. It turned out that details reflect hidden time series features which come in handy for near-term and medium-term social processes forecasting.
- We can use the fast Mallat's pyramid algorithm [10]. Its runtime complexity is $O(n)$, where $n$ is the maximum wavelet decomposition level.
- WT is already being successfully and intensively used to provide individual anonymity [11].

Thereby, in this work we set and solve a following task. We want to provide group anonymity for depersonalized respondent data according to particular attribute combination. We propose to complete this task using WT. In this case, group anonymity is gained through redistributing wavelet approximation values. Fixing data mean value and leaving wavelet details unchanged (or proportionally altering them) preserves data features which might become useful for specific researches. Figuratively speaking, we change the relief (approximation) of a restricted area, but try to preserve local data distribution (details).

We would also like to admit that there is no feasible algorithm to restore primary data after modifying them using the proposed method.

## 2 Theoretic Background

### 2.1 General Group Anonymity Definitions

Let the microfile data be presented as Table 1. In this table, $\mu$ stands for the number of records (respondents), $\eta$ stands for the number of attributes, $r_i$ stands for the $i^{th}$ record, $u_j$ stands for the $j^{th}$ attribute, $z_{ij}$ stands for a microfile data element.

**Table 1.** Microfile data.

|       | $u_1$      | $u_2$      | ...   | $u_\eta$      |
|-------|------------|------------|-------|---------------|
| $r_1$ | $z_{11}$   | $z_{12}$   | ...   | $z_{1\eta}$   |
| $r_2$ | $z_{21}$   | $z_{22}$   | ...   | $z_{2\eta}$   |
| ...   | ...        | ...        | ...   | ...           |
| $r_\mu$ | $z_{\mu 1}$ | $z_{\mu 2}$ | ... | $z_{\mu\eta}$ |

To provide group anonymity we need to decide first which attribute values and of what groups we would like to protect.

Let us denote by $S_v$ a subset of a Cartesian product $u_{v_1} \times u_{v_2} \times ... \times u_{v_l}$ of Table 1 columns. Here, $v_i$, $i = \overline{1,l}$ are integers. We will call an element $s_k^{(v)} \in S_v$, $k = \overline{1, l_v}$, $l_v \leq \mu$ *a vital value combination* because such combinations are vital for solving our task. Respectively, each element of $s_k^{(v)}$ will be called *a vital value*, and $u_{v_j}$ will be called *a vital attribute*.

Our task is to protect some of the vital value combinations. E.g., if we took "Age" and "Employment status" as vital attributes we could possibly be interested in providing anonymity for the vital value combination ("Middle-aged"; "Unemployed").

We will also denote by $S_p$ a subset of microfile data elements corresponding to the $p^{th}$ attribute, $p \neq v_i$ $\forall i = \overline{1,l}$. Elements $s_k^{(p)} \in S_p$, $k = \overline{1, l_p}$, $l_p \leq \mu$ will be called *parameter values*, whereas $p^{th}$ attribute will be called *a parameter attribute* because it will be used for dividing microfile data into groups to be analyzed.

For example, if we took "Place of living" as a parameter attribute we could obtain groups of "Urban" and "Rural" residents.

After having defined both parameter and vital attributes and values, we need to calculate the quantities of respondents that correspond to a specific pair of a vital value combination and a parameter value

These quantities can be gathered into an array $q = (q_1, q_2, ..., q_m)$ which we will call *a quantity signal*.

To provide group anonymity for the microfile we need to replace this quantity signal with another one: $\tilde{q} = (\tilde{q}_1, \tilde{q}_2, ..., \tilde{q}_m)$. Also, we need to preserve specific data features.

First of all, we need to make sure that the overall number of records remains stable:
$$\sum_{i=1}^{m} q_i = \sum_{i=1}^{m} \tilde{q}_i .$$

And, as it was mentioned in Section 1, we also need to preserve all the wavelet decomposition details of signal $q$ up to some level $k$ (or at least alter them proportionally).

Possible solution to the task is proposed in the following subsections.

## 2.2 General Wavelet Transform Definitions

In this subsection we will revise the WT basics which are necessary for the further explanation. For detailed information see [10].

Let us call an array $s = (s_1, s_2, ..., s_m)$ of discrete values a signal.

Let a high-pass wavelet filter be denoted as $h = (h_1, h_2, ..., h_n)$, and a low-pass wavelet filter be denoted as $l = (l_1, l_2, ..., l_n)$.

To perform signal $s$ one-level wavelet decomposition, we need to carry out following operations:

$$a_1 = s *_{\downarrow 2n} l; \quad d_1 = s *_{\downarrow 2n} h. \quad (1)$$

In **(1)**, a convolution (which is denoted by $*$) of $s$ and $l$ is taken, and then the result is being dyadically downsampled (denoted by $_{\downarrow 2n}$). Also, $a_1$ is an array of approximation coefficients, whereas $d_1$ is an array of detail coefficients.

To obtain approximation and detail coefficients at level $k$, we need to perform **(1)** on approximation coefficients at level $k-1$:

$$a_k = a_{k-1} *_{\downarrow 2n} l = ((s \underbrace{*_{\downarrow 2n} l) ... *_{\downarrow 2n}}_{k \text{ times}} l); \quad d_k = a_{k-1} *_{\downarrow 2n} h = (((s \underbrace{*_{\downarrow 2n} l) ... *_{\downarrow 2n}}_{k-1 \text{ times}} l) *_{\downarrow 2n} h). \quad (2)$$

We can always present an initial signal $s$ as

$$s = A_k + \sum_{i=1}^{k} D_i. \quad (3)$$

Here, $A_k$ is called an approximation at level $k$, and $D_i$ is called a detail at level $i$. Approximation and details from **(3)** can be presented as follows:

$$A_k = ((a_k \underbrace{*_{\uparrow 2n} l) ... *_{\uparrow 2n}}_{k \text{ times}} l); \quad D_k = (((d_k *_{\uparrow 2n} h) \underbrace{*_{\uparrow 2n} l) ... *_{\uparrow 2n}}_{k\text{-}1 \text{ times}} l). \quad (4)$$

In **(4)**, $a_k$ and $d_k$ are being dyadically upsampled (which is denoted by $_{\uparrow 2n}$) first, and then convoluted with the appropriate wavelet filter.

As we can see, all $A_k$ elements depend on the $a_k$ coefficients.

According to Section 1, we need to somehow modify the approximation, and at the same time preserve the details. As it follows from **(4)**, details do not depend on approximation coefficients. Thus, preserving detail coefficients preserves the details.

Respectively, to modify the approximation we have to modify corresponding approximation coefficients.

## 2.3 Obtaining New Approximation Using Wavelet Reconstruction Matrices

In [12], it is shown that WT can be performed using matrix multiplications. In particular, we can always construct a matrix such that

$$A_k = M_{rec} \cdot a_k. \tag{5}$$

For example, $M_{rec}$ can be obtained by consequent multiplication of appropriate upsampling and convolution matrices.

We will call $M_{rec}$ a wavelet reconstruction matrix (WRM).

Now, let us apply WRM to solve the problem stated in Subsection 2.1.

Let $q = (q_1, q_2, ..., q_m)$ be a quantity signal of length $m$. Let also $l = (l_1, l_2, ..., l_n)$ denote a low-pass wavelet filter.

Taking into consideration **(5)**, all we need to do is to find new coefficients $\widehat{a}_k$. For example, they can be found by solving a linear programming problem with constraints obtained from matrix $M_{rec}$. Then, adding new approximation $\widehat{A}_k$ and all the details corresponding to $q$, we can get a new quantity signal $\widehat{q}$.

In many cases, adding $\widehat{A}_k$ can result in the negative values of a new signal $\widehat{q}$, which is totally unacceptable. In this case we can modify $\widehat{q}$ to make it non-negative (e.g., by adding to each element of $\widehat{q}$ a suitable value), and thus receive a new signal $\hat{q}$.

Another problem arises. The mean value of the resultant signal $\hat{q}$ will obviously differ from the initial one. To overcome this problem, we need to multiply it by such a coefficient that the result has the required mean value. Due to the algebraic properties of convolution, both resultant details' and approximation' absolute values will differ from the initial ones by that precise coefficient. This means that the details will be changed proportionally which totally suits our problem statement requirements. In result, we obtain our required signal $\tilde{q}$.

To illustrate this method we will consider a practical example.

## 3   Experimental Results

To show the method under review in action, we took the 5-Percent Public Use Microdata Sample Files from U.S. Census Bureau [13] corresponding to the 2000 U.S. Census microfile data on the state of California.

The microfile provides various information on more than 1,6 million respondents. We took a "Military service" attribute as a vital one. This attribute is a categorical one. Its values are integers from 0 to 4. For simplicity, we took one vital value combination consisting of only one vital value, i.e. "1". It stands for "Active duty".

We also took "Place of Work Super-PUMA" as a parameter attribute. This attribute is also a categorical one. Its values stand for different statistical area codes.

For our example, we decided to take the following attribute values as parameter ones: 06010, 06020, 06030, 06040, 06060, 06070, 06080, 06090, 06130, 06170, 06200, 06220, 06230, 06409, 06600 and 06700. These codes correspond to border, coastal and island statistical areas.

By choosing these exact attributes we actually set a task of protecting information on military officers' number distribution over particular Californian statistical areas.

According to Section 2, we need to construct an appropriate quantity signal. The simplest way to do that is to count respondents with appropriate pair of a vital value combination and a parameter value. The results are shown in Table 2 (the third row).

Let's use the second order Daubechies low-pass wavelet filter $l \equiv \left( \dfrac{1+\sqrt{3}}{4\sqrt{2}}, \dfrac{3+\sqrt{3}}{4\sqrt{2}}, \dfrac{3-\sqrt{3}}{4\sqrt{2}}, \dfrac{1-\sqrt{3}}{4\sqrt{2}} \right)$ to perform two-level wavelet decomposition (**2**) of a corresponding quantity signal (all the calculations were carried out with 12 decimal numbers, but we will present all the numeric data with 3 decimal numbers):

$a_2 = (a_2(1), a_2(2), a_2(3), a_2(4)) = (2272.128, 136.352, 158.422, 569.098)$.

Now, let us construct a suitable WRM:

$$M_{rec} = \begin{pmatrix} 0.637 & 0 & 0 & -0.137 \\ 0.296 & 0.233 & 0 & -0.029 \\ 0.079 & 0.404 & 0 & 0.017 \\ -0.012 & 0.512 & 0 & 0 \\ -0.137 & 0.637 & 0 & 0 \\ -0.029 & 0.296 & 0.233 & 0 \\ 0.017 & 0.079 & 0.404 & 0 \\ 0 & -0.012 & 0.512 & 0 \\ 0 & -0.137 & 0.637 & 0 \\ 0 & -0.029 & 0.296 & 0.233 \\ 0 & 0.017 & 0.079 & 0.404 \\ 0 & 0 & -0.012 & 0.512 \\ 0 & 0 & -0.137 & 0.637 \\ 0.233 & 0 & -0.029 & 0.296 \\ 0.404 & 0 & 0.017 & 0.079 \\ 0.512 & 0 & 0 & -0.012 \end{pmatrix}.$$

According to (**5**), we obtain a signal approximation:

$A_2 = (1369.821, 687.286, 244.677, 41.992, -224.98, 11.373, 112.86, 79.481, 82.24, 175.643, 244.757, 289.584, 340.918, 693.698, 965.706, 1156.942)$.

As we can see, according to the extremum transition approach, we have to lower the military men quantity in the 06700 area. At the same time, we have to raise appropriate quantities in some other areas. The particular choice either may depend on any additional goals to achieve or it can be absolutely arbitrary.

But, along with this, we have to avoid incidental raising of the other signal elements. We can achieve this by using appropriate constraints. Also, it is necessary to note down that there can possibly be some signal elements which do not play important role, i.e. we can change them without any restrictions.

To show how to formally express suitable constraints, we decided to raise the quantities in such central-part signal elements like 06070, 06080, 06090, 06130, 06170 and 06200; besides, we have chosen the first and the last three signal elements to lower their values.

Considering these requirements, we get the following constraints:

$$\begin{cases} 0.637 \cdot \hat{a}_2(1) - 0.137 \cdot \hat{a}_2(4) \leq 1369.821 \\ 0.296 \cdot \hat{a}_2(1) + 0.233 \cdot \hat{a}_2(2) - 0.029 \cdot \hat{a}_2(4) \leq 687.286 \\ 0.079 \cdot \hat{a}_2(1) + 0.404 \cdot \hat{a}_2(2) + 0.017 \cdot \hat{a}_2(4) \leq 244.677 \\ -0.137 \cdot \hat{a}_2(1) + 0.637 \cdot \hat{a}_2(2) \geq -224.980 \\ -0.029 \cdot \hat{a}_2(1) + 0.296 \cdot \hat{a}_2(2) + 0.233 \cdot \hat{a}_2(3) \geq 11.373 \\ 0.017 \cdot \hat{a}_2(1) + 0.079 \cdot \hat{a}_2(2) + 0.404 \cdot \hat{a}_2(3) \geq 112.860 \\ -0.012 \cdot \hat{a}_2(2) + 0.512 \cdot \hat{a}_2(3) \geq 79.481 \\ -0.137 \cdot \hat{a}_2(2) + 0.637 \cdot \hat{a}_2(3) \geq 82.240 \\ -0.029 \cdot \hat{a}_2(2) + 0.296 \cdot \hat{a}_2(3) + 0.233 \cdot \hat{a}_2(4) \geq 175.643 \\ 0.233 \cdot \hat{a}_2(1) - 0.029 \cdot \hat{a}_2(3) + 0.296 \cdot \hat{a}_2(4) \leq 693.698 \\ 0.404 \cdot \hat{a}_2(1) + 0.017 \cdot \hat{a}_2(3) + 0.079 \cdot \hat{a}_2(4) \leq 965.706 \\ 0.512 \cdot \hat{a}_2(1) - 0.012 \cdot \hat{a}_2(4) \leq 1156.942 \,. \end{cases}$$

A possible solution is $\hat{a}_2 = (0, 379.097, 31805.084, 5464.854)$.

Using $M_{rec}$ and **(5)**, we can get a new approximation:

$\hat{A}_2 = (-750.103, -70.090, 244.677, 194.196, 241.583, 7530.756, 12879.498, 16287.810, 20216.058, 10670.153, 4734.636, 2409.508, -883.021, 693.698, 965.706, -66.997)$.

Since our integral aim is to preserve signal details, we construct our masked quantity signal by adding a new approximation and primary details:

$\hat{q} = \hat{A}_2 + D_1 + D_2 = (-2100.924, -745.376, 153.000, 223.204, 479.563, 7598.383, 12773.639, 16241.328, 20149.818, 10764.510, 5301.879, 2254.924, -982.939, 14.000, 60.000, 3113.061)$.

As we can see, some signal elements are negative. Since quantities cannot be negative, we need to add to every signal's element an appropriate value, e.g. 2500:

$\hat{q} = (399.076, 1754.624, 2653.000, 2723.204, 2979.563, 10098.383, 15273.639, 18741.328, 22649.818, 13264.510, 7801.879, 4754.924, 1517.061, 2514.000, 2560.000, 5613.061)$.

Here, all the signal samples are non-negative. Therefore, the only requirement not fulfilled yet is the equality of corresponding mean values. To provide that, we need to multiply $\hat{q}$ by the coefficient $\sum_{i=1}^{16} q_i / \sum_{i=1}^{16} \hat{q}_i = 0.054$.

The resultant signal has the same mean value and wavelet decomposition details as the initial one. This can be checked-up through easy but rather cumbersome calculations.

Since quantities can be only integers, we need to round the signal. Finally, we get the required quantity signal $\tilde{q}$ (see Table 2, the fourth row).

As we can see, the masked data are completely different from the primary ones, though both mean value and wavelet decomposition details are preserved.

To finish the task, we need to compile a new microfile. It is always possible to do as long as there are enough records to modify vital values of. Anyway, we can always demand this when building-up linear programming problem constraints.

**Table 2.** Quantity signals for the U.S. Census Bureau microfile.

| Column number | 1 | 2 | 3 | 4 | 5 | 6 | 7 | 8 |
|---|---|---|---|---|---|---|---|---|
| Area code | 06010 | 06020 | 06030 | 06040 | 06060 | 06070 | 06080 | 06090 |
| Signal $q$ | 19 | 12 | 153 | 71 | 13 | 79 | 7 | 33 |
| Signal $\tilde{q}$ | 22 | 95 | 144 | 148 | 162 | 549 | 831 | 1019 |

| Column number | 9 | 10 | 11 | 12 | 13 | 14 | 15 | 16 |
|---|---|---|---|---|---|---|---|---|
| Area code | 06130 | 06170 | 06200 | 06220 | 06230 | 06409 | 06600 | 06700 |
| Signal $q$ | 16 | 270 | 812 | 135 | 241 | 14 | 60 | 4337 |
| Signal $\tilde{q}$ | 1232 | 722 | 424 | 259 | 83 | 137 | 139 | 305 |

## 4  Conclusion and Future Research

In the paper, we have set the task of providing group anonymity as a task of protecting such collective data patterns that cannot be retrieved by analyzing individual information only.

We have proposed a wavelet-based method which aims at preserving the data wavelet details as a source of information on the data patterns and relations between their components with different frequencies, along with the data mean value. At the same time, the method actually provides group anonymity since an appropriate level of uncertainty is being introduced into the data (by modifying the wavelet approximation).

The method is relatively easy and can be implemented programatically. Also, the method is rather flexible and can yield various resultant data sets depending on the particular task definition. Moreover, it can be combined with any existing individual anonymity methods to gain the most efficiently protected datasets.

On the other hand, the method isn't acceptable in various cases because it doesn't guarantee that some statistical data features, such as standard deviation, persist.

In the paper, we only pointed out the problem of group anonymity. There remain many questions to answers and challenges to response. Among them we would especially like to stress on such ones:

- Using different wavelet bases can lead to obtaining different data sets.
- Modifying quantity signals isn't very useful for different real-life examples. In situations like protecting the regional distribution of middle-aged people the relative data such as ratios seem to be more important to protect.

- In general, it is not always easy to define parameter and vital sets to determine the records to redistribute. This procedure also needs to be studied thoroughly in the future.